\documentclass[twocolumn,showpacs,preprintnumbers,amsmath,amssymb]{revtex4}

\usepackage{graphicx}
\usepackage{dcolumn}
\usepackage{bm}

\begin{document}
\title{Single-band tight-binding parameters for Fe-MgO-Fe magnetic heterostructures}

\author{Tehseen Z. Raza}
\address{School of Electrical and Computer Engineering and NSF Network for Computational Nanotechnology,
Purdue University, West Lafayette, IN 47907}
\author{Hassan Raza}
\address{School of Electrical and Computer Engineering, Cornell University, Ithaca, NY 14853}

\begin{abstract}
We present a computationally efficient transferable single-band tight-binding model (SBTB) for spin polarized transport in heterostructures with an effort to capture the band structure effects. As an example, we apply it to study transport through Fe-MgO-Fe(100) magnetic tunnel junction devices. We propose a novel approach to extract suitable tight-binding parameters for a material by using the energy resolved transmission as the benchmark, which inherently has the bandstructure effects over the two dimensional transverse Brillouin zone. The SBTB parameters for each of the four symmetry bands for bcc Fe(100) are first proposed which are complemented with the transferable tight-binding parameters for the MgO tunnel barrier for the $\Delta_1$ and $\Delta_5$ bands. The non-equilibrium Green's function formalism is then used to calculate the transport. Features like I-V characteristics, voltage dependence and the barrier width dependence of the tunnel magnetoresistance ratio are captured quantitatively and the trends match well with the ones observed by \textit{ab initio} methods.
\end{abstract}

\maketitle

\section{Introduction}

Transport across multilayered heterostructures is a problem of great interest both for its intrinsic physics as well as its device applications \cite{Magnetic_Book}. Very often each of the component materials has been studied extensively on its own, but it is difficult to make use of this knowledge because various studies employ different models and it is difficult to combine their results. As a result, the only viable approach is to start anew for each heterostructure. The objective of this paper is to present a scheme for extracting suitable single-band tight-binding (SBTB) parameters for each of the component materials, thus translating the results of earlier studies all into one common SBTB platform which can then be used in a standard non-equilibrium Green's function (NEGF) based model for quantum transport. We illustrate our approach with an example of great current interest, namely an Fe-MgO-Fe magnetic tunnel junction (MTJ) device. We use the principles described in this paper to obtain the SBTB parameters for bcc Fe(100) from the extended H\"uckel parameters \cite{Cerda00} and those for MgO from \textit{ab initio} models \cite{Heiliger05, Heiliger06, Heiliger08}. Using these SBTB parameters, extracted from different sources, in an NEGF model for transport we obtain I-V characteristics, voltage dependence of tunnel magnetoresistance (TMR) ratios and barrier width dependence of TMR in good agreement with published first-principles results for the same structure.

\begin{figure}
\centering
\includegraphics[width=3.4in]{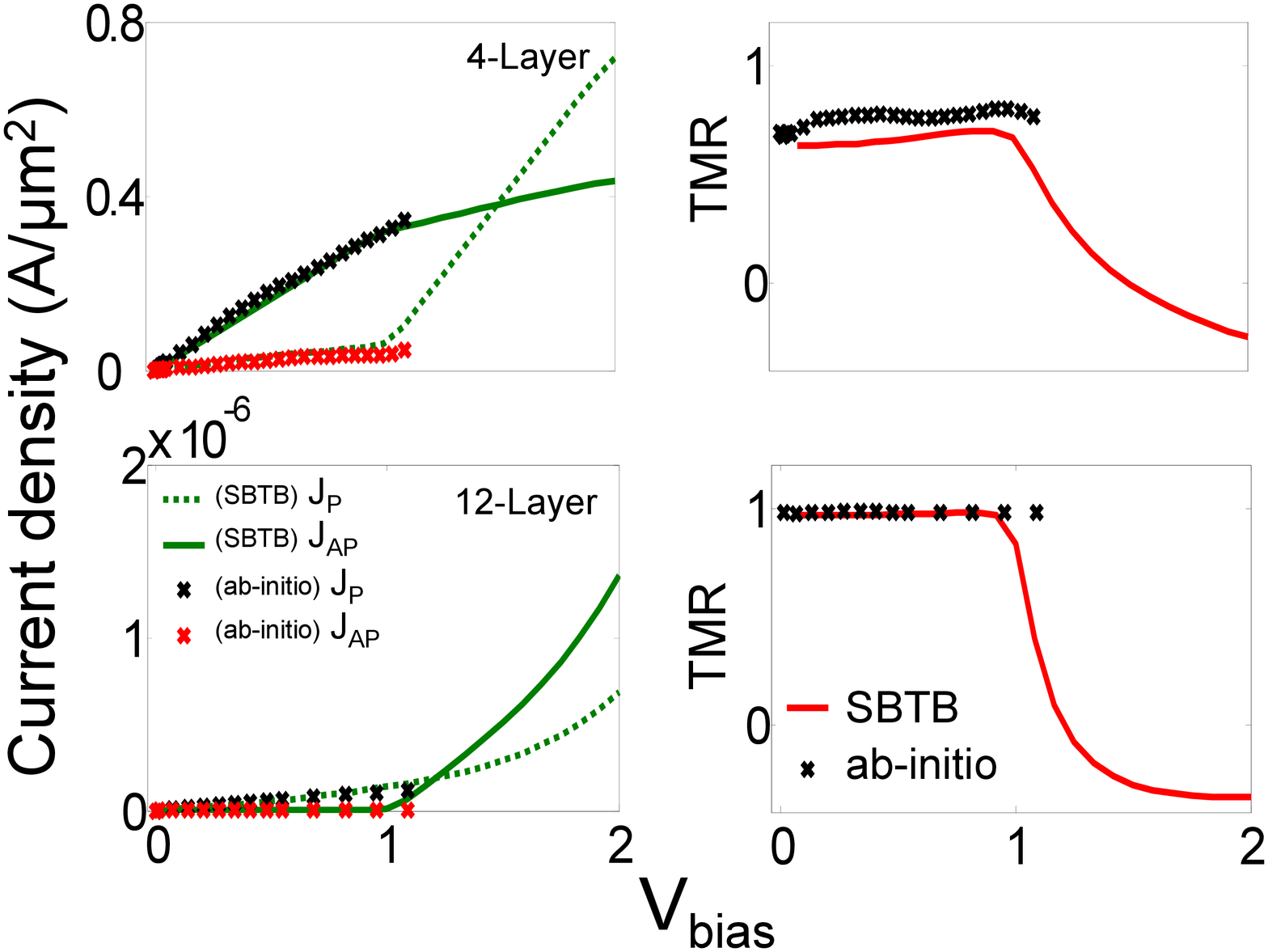}
\caption{Total current densities for the parallel and the anti-parallel configurations and the tunnel magneto resistance ratio (TMR) for a 4-layer (top) and 12-layer (bottom) device. SBTB transferable parameters are optimized for the MgO tunnel barrier to match the current levels from the \textit{ab initio} calculations \cite{Heiliger05, Heiliger06, Heiliger08}. The bias dependence of TMR is also captured well within this simple single-band tight-binding model.}
\end{figure}

\textit{Ab initio} modeling of materials is at an advanced stage \cite{Martin04}. Coupled with the quantum transport, these models have been successfully applied to nanoscale systems and heterostructures. However, such methods are resource intensive, and simplified models that capture the essential physics due to the underlying electronic structure effects are desirable. Since the seminal work of studying material properties using tight binding parameters by Slater and Koster \cite{Slater54}, there has been a motivation to develop computationally efficient yet accurate models to capture the underlying physical mechanisms and the bandstructure effects. Such simple methods may or may not capture all the intricate details present in the more sophisticated models, \textit{e.g.} \textit{ab initio}, semi-empirical tight-binding \cite{HassanSi, HassanacGNR} and empirical tight-binding methods, but they do provide a platform for large scale calculations. 

The simplest method currently available in this context is an effective mass model. Although very successful for matching experiments, it is well known that it does not capture the electronic structure effects. We find the same for bcc Fe(100) because the band dispersions do not remain parabolic over the two dimensional (2D) transverse Brillouin zone (BZ). Furthermore, the energy bands have finite width, whereas bandwidth is infinite in continuum effective mass models and it depends on the lattice spacing in discrete effective mass models. Moreover, the various band edges shift over the 2D BZ and this shift is not necessarily dictated by the transverse mode energy $\hbar^2k_t^2/2m_t$, as prescribed by an effective mass model. Our objective in this paper is to propose a computationally efficient method for heterostructures with an effort to incorporate the physics of electronic structure effects over the 2D transverse BZ in the transport quantities. Motivated by this, we propose a single-band tight-binding (SBTB) model for transport through heterostructures and outline the procedure for extracting suitable tight-binding parameters. The computational complexity of this model is the same as that of an effective mass model. 

Recently, Fe-MgO-Fe heterostructures have appeared as the most noteworthy example in spintronics \cite{Ikeda07}. These devices have emerged as one of the candidates for random access memory applications. The prediction of high tunnel magneto resistance (TMR) ratio for crystalline MgO barrier of over 1000$\%$ \cite{Butler01, Mathon01} was followed by observations of about 200$\%$ TMR ratios at room temperature in Fe-MgO-Fe and CoFe-MgO-CoFe MTJ devices \cite{Yuasa04,Parkin04}. Since then, there has been an increased effort to integrate them into practical devices. Although, \textit{ab initio} \cite{Butler01, Heiliger05, Heiliger06, Heiliger08, Zhang04, Derek07} and empirical tight binding \cite{Mathon01} studies have been reported, their computational complexity limit their use for rapid device prototyping. This method not only allows simulating device characteristics efficiently but it also gives an inherently simple and intuitive understanding for the underlying device physics. A comparison between the SBTB and ab initio models is shown in Fig. 1. For low bias, it matches well with Ref. \cite{Heiliger08}. At high bias, our model predicts TMR roll-off and ultimately becoming negative, which was later observed in Ref. \cite{Ivan08}. 

\begin{figure}
\centering
\includegraphics[width=3.25in]{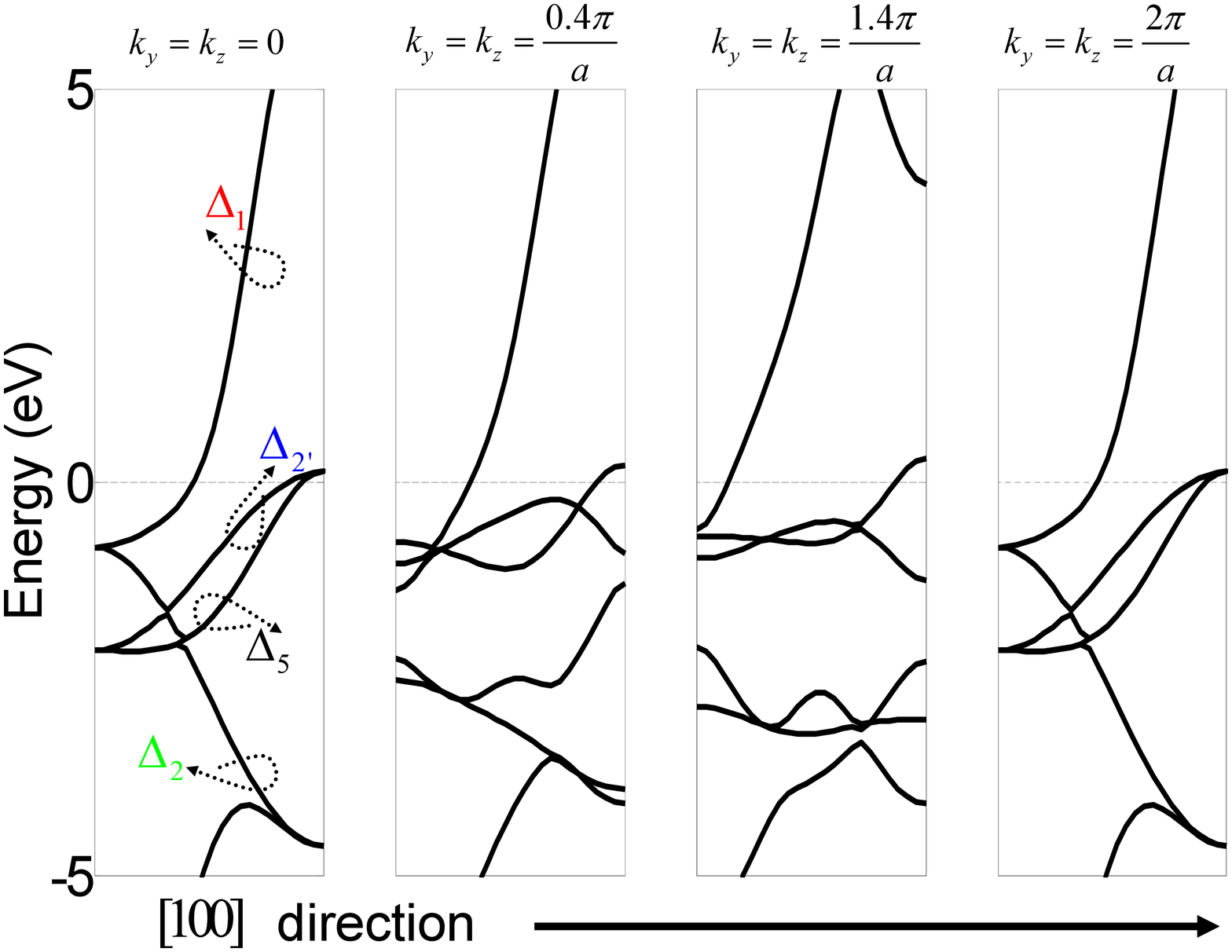}
\caption{Band structure for the majority spin over the two dimensional transverse Brillouin zone computed using EHT. In general, the bands are not parabolic and hence an effective mass description is not valid.}
\end{figure}

\begin{figure}
\centering
\includegraphics[width=2.6in]{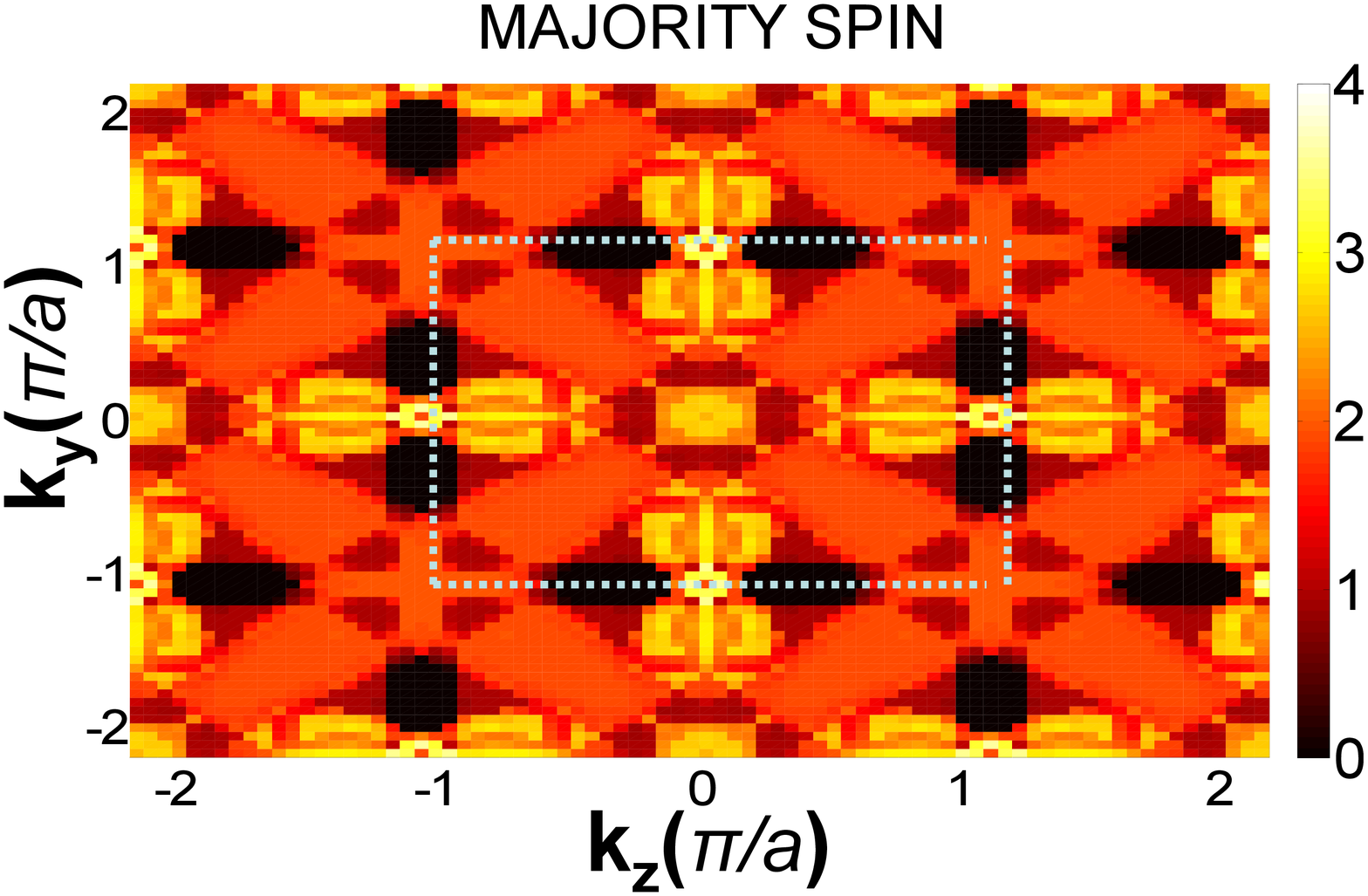}
\includegraphics[width=2.6in]{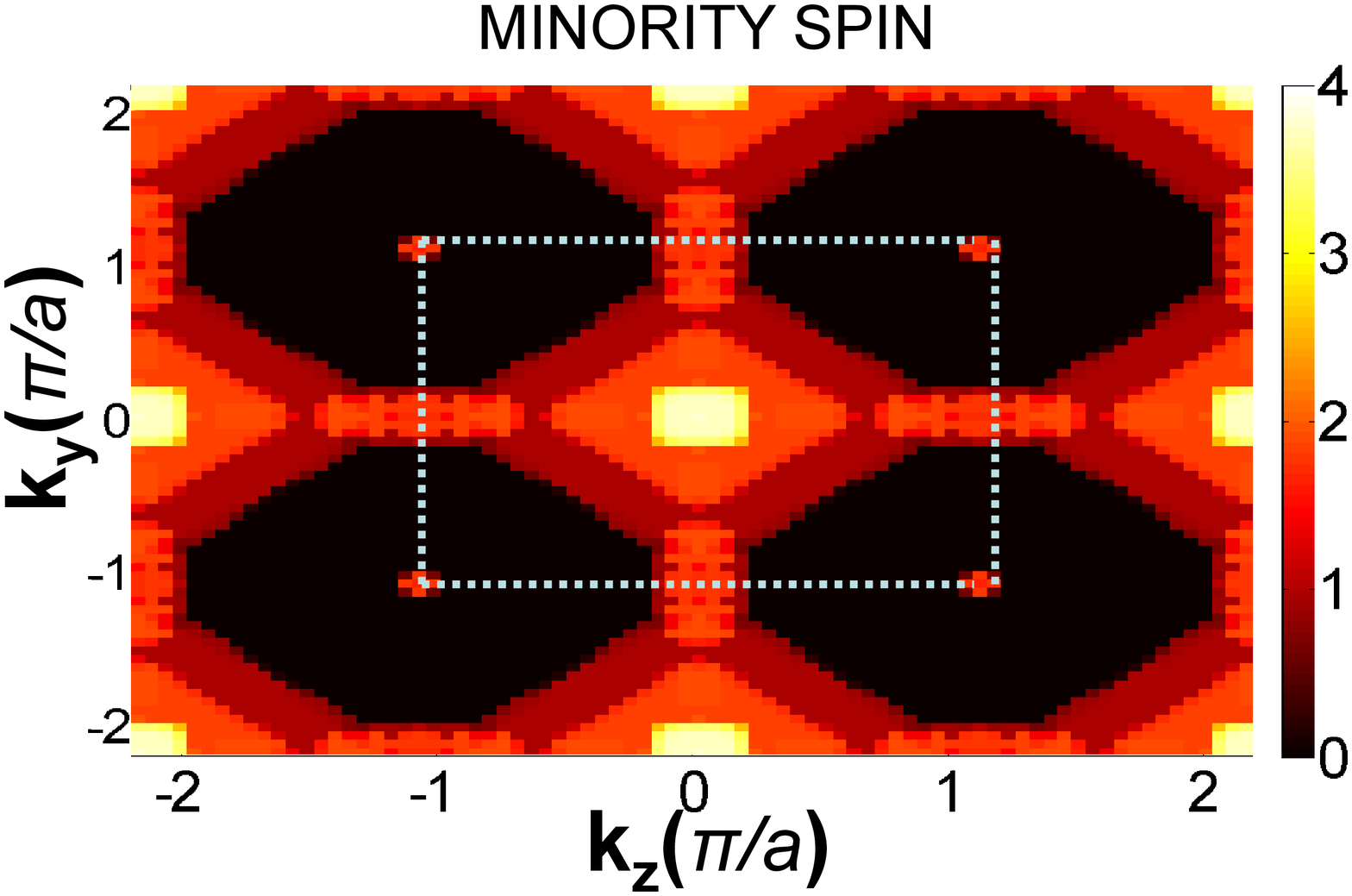}
\caption{Equilibrium transmission over the two dimensional transverse BZ for $E_f=0$. $k_{||}=(k_{x},k_{y})$ resolved transmission for the majority and the minority spin. The BZ is shown by the dotted line. The transmission has units of $10^{19}\ m^{-2}$ and $k$ has units of $m^{-2}$.}
\end{figure}

Our approach is to present SBTB parameters for different materials independently and then couple them across the interface. Instead of fitting the parabolic bands to the bandstructure to extract effective masses, we propose to match the energy resolved transmission. The energy resolved transmission plots reflect the band structure effects for the majority and the minority spin bands over the transverse BZ and we propose to capture this effect in an average manner with the SBTB parameters. As shown in Fig. 3, the transmission is strongly dependent on the transverse BZ, and it is imperative to incorporate these effects. Thus, we first extract the SBTB parameters for bcc Fe(100) from the energy resolved transmission plots calculated using a semi-empirical atomistic method based on extended H\"uckel theory (EHT). Next, we propose the barrier parameters, for tunneling through the MgO region for the $\Delta_1$ and $\Delta_5$ bands by comparing with the I-V characteristics through Fe-MgO-Fe calculated using \textit{ab initio} methods \cite{Heiliger05, Heiliger06, Heiliger08}. Only $\Delta_1$ and $\Delta_5$ bands are considered and $\Delta_2$ and $\Delta_{2'}$ bands are ignored due to their large decay rates \cite{Butler01}. 

This paper is divided into four sections. In Sec. II, we discuss the theoretical approach and the assumptions made. In Sec. III, we discuss the results. Finally, in Sec. IV, we provide the conclusions. 

\begin{table}[!t]
\renewcommand{\arraystretch}{1.3}
\caption{\label{tab:Table1} SBTB parameters for the majority($\uparrow$) and the minority($\downarrow$) spin bands of bcc Fe(100). The band offsets ($E_{bo}$) for $\Delta_{1}$, $\Delta_{2}$, $\Delta_{2}'$ and $\Delta_{5}$ bands are shown. Orbital symmetry states of bands are also given \cite{Tiusan07}.}
\begin{tabular}{llcccc}
\hline\hline
Band          &Symmetry                             &$t_{o}(eV)$&         & $E_{bo}(eV)$\\ \hline
              &                                     &$\uparrow$  &  $\downarrow$       &  $\uparrow$  &  $\downarrow$ \\
\hline\hline
$\Delta_{1}$  & $4s, 4p_{z}, 3d_{z^{2}}$            & 2.5         &  2.5                & -1             & 1\\
$\Delta_{2'}$ & $3d_{xy}$                           & 0.2         &  0.2                & -1.5           & 0.4\\
$\Delta_{5}$  & $4p_{x}, 4p_{y}, 3d_{xz}, 3d_{yz}$  & 1           &  1                  & -3.5           & -2.0\\
$\Delta_{2}$  & $3d_{x^{2}-y^{2}}$                  & -0.2        &-0.35                & -2.1           & -0.8\\
\hline
\end{tabular}
\end{table} 

\begin{table}[!t]
\renewcommand{\arraystretch}{1.3}
\caption{SBTB parameters for MgO. $U_{b}$ is the MgO barrier height, $t_{o}$ is the hopping parameter for $\Delta_1$ and $\Delta_5$ bands.}
\label{tab:Table2}
\centering
\begin{tabular}{lcc}
\hline\hline
               & $t_{o}(eV)$                 &  $U_{b}(eV)$         \\
\hline\hline
$\Delta_{1}$ band             & 0.64                    &  2.8             \\
$\Delta_{5}$ band             & 0.64                    &  4.5             \\
\hline
\end{tabular}
\end{table}

\section{Theoretical model and assumptions} 
To calculate the transmission for homogeneous materials using the SBTB method, for each band we start with the following Hamiltonian \cite{Kittel_Book},
\begin{eqnarray}H_{SBTB}=\begin{cases}E_{bo}+2t_o\ \ \ for\ i=j\cr -t_o\ \ \ \ \ \ \ \ \ \ for\ |i-j|=1 \end{cases}\end{eqnarray}
where $E_{bo}$ is the band offset and $t_o$ is the hopping parameter. This results in cosine dispersion as:
\begin{eqnarray}\epsilon(k)=E_{bo}+2t_{o}[1-\cos(ka)]\end{eqnarray} 
which gives a bandwidth of $4t_o$. Here $a$ is the lattice spacing and $k$ is the wave vector in the transport direction, where $a=4.2\AA$ and $2.86\AA$ for MgO and Fe respectively. This dispersion is exactly the same as that of an effective mass Hamiltonian discretized using the finite-difference method. However, in SBTB, $t_{o}$ is a constant whereas in effective mass approach it is inversely proportional to the lattice spacing. In SBTB method, each lattice point corresponds to a unit cell. Therefore, corresponding to a ten layer (five unit cells) device considered to calculate transmission using the EHT calculations, there are five lattice points in the SBTB model. The extracted SBTB parameters ($E_{bo}$ and $t_o$) for various bcc Fe(100) symmetry bands are shown in Table I. 

For heterostructures, the above Hamiltonian is modified depending on the device material. \textit{e.g.}, in our system the device is made up of a tunnel barrier, hence $U_L$ is the Laplace potential linearly dropped across the insulator region, which is added to the MgO Hamiltonian matrix as follows: 
\begin{eqnarray} H_{SBTB}=\begin{cases}E_{bo}+2t_o+U_L(i,j)\ for\ i=j\cr -t_o\ \ \ \ \ \ \ \ \ \ \ \ \ \ \ \ \ \ \ \ \ \ for\ |i-j|=1\end{cases}\end{eqnarray}
For MgO, the hopping parameters ($t_o$) and the band offsets ($E_{bo} = E_f+U_b$) are given in Table II. The method of extracting these band parameters will be explained later in this section. 

At the heterostructure interface, the off-diagonal elements are taken such that the resulting Hamiltonian is Hermitian to ensure that the energy eigenvalues are real and current is conserved \cite{Datta_Book}. Additionally, the effect of different interface structures may be incorporated by introducing additional SBTB parameter for the interface, and is left for future work. 

We then use the equilibrium part of NEGF \cite{Datta_Book} to calculate the transmission ($\hat{T}$), which gives unity transmission in the bandwidth region due to the one-dimensional nature of the transport. Finally, the transmission per unit area is calculated analytically by summing over the 2D transverse BZ as:
\begin{eqnarray}T_{SBTB}=\frac{1}{4\pi^2}\int_{(-\frac{\pi}{a},-\frac{\pi}{a})}^{(\frac{\pi}{a},\frac{\pi}{a})}dk_{||}\hat{T}(E_l)=\frac{1}{a^2}\hat{T}\end{eqnarray} 
where $a=2.86\AA$ is the cubic Fe lattice constant. This equation is the key equation for the SBTB method, where the energy resolved transmission can be captured averagely by a suitable band width for a band specified by a single hopping parameter $t_{o}$ and the band offset $E_{bo}$. 

\begin{figure}
\centering
\includegraphics[width=3.4in]{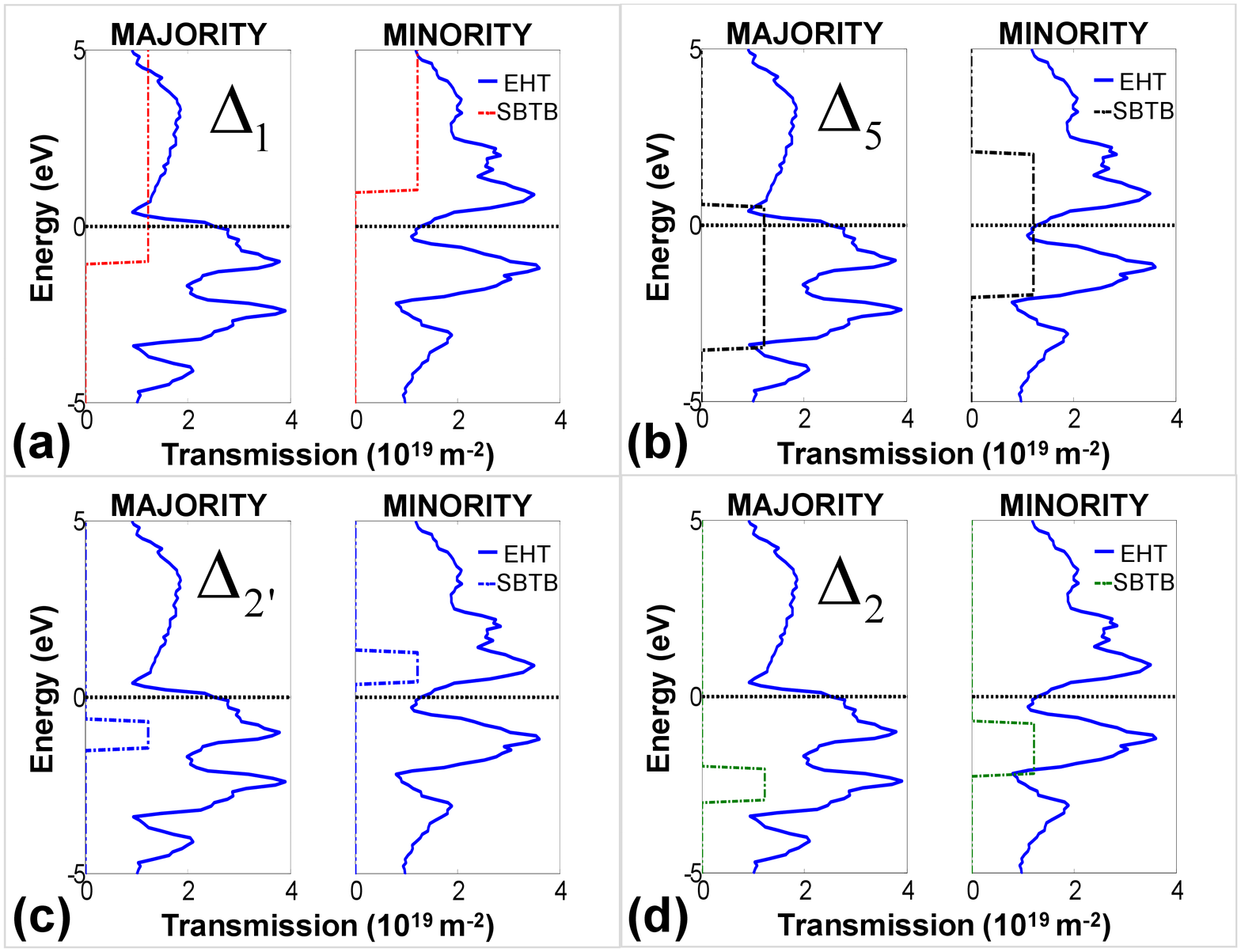}
\caption{The proposed individual band contribution to the energy resolved transmission per unit area using SBTB. Transmission per unit area through bulk Fe using EHT (solid line) and for $\Delta_1$, $\Delta_{2}$, $\Delta_{2'}$ and $\Delta_{5}$ bands using SBTB (dashed line) for the majority (left) and the minority (right) spin. The SBTB transferable parameters for these bands are estimated from their finite bandwidths in the energy resolved transmission and are shown in Table I.}
\end{figure}

We now present the method of extracting the suitable band parameters for bcc Fe(100) and MgO. Due to the 1D nature of the transport for the system under study, the transmission is unity within the bandwidth. Therefore, instead of matching the fine details in the transmission as shown in Fig. 3 due to the varying band offsets and bandwidths over the 2D transverse BZ, we capture the transmission in an average way by estimating the bandwidths and band offsets for the various symmetry bands. Unlike an effective mass model with infinite bandwidth, the finite bandwidth in SBTB model is important to match the transmission. Since this bandwidth equals $4t_o$ (see eq. 2), the hopping parameters $t_o$ and band offsets $E_{bo}$ can then be calculated. 

The transmission calculated for bcc Fe(100) using the EHT method serves as a benchmark for the SBTB calculations. Adapted from Ref. \cite{Tiusan07} and also shown in Table II, $\Delta_2$ and $\Delta_{2'}$ bands have d-type orbital symmetry. At $k_{||}=0$, the $\Delta_5$ band is doubly degenerate. However for $k_{||} \neq 0$, this degeneracy is lifted. It is well established that the bands with d-type orbital symmetry are localized in energy and have smaller bandwidths due to reduced hybridization. We therefore attribute the two peaks in transmission due to the $\Delta_{2'}$ and $\Delta_{2}$ symmetry bands superimposed on the $\Delta_{1}$ and $\Delta_{5}$ bands as shown in Fig. 4. We estimate the average bandwidths and band offsets from the above-mentioned analysis and subsequently extract $t_o$. The transmissions for $\Delta_1$, $\Delta_2$, $\Delta_{2'}$ and $\Delta_5$ bands are shown in Fig. 4. More sophisticated methods may be used for estimating the bandwidths, which is left for future work.

For MgO parameterization, current through Fe-MgO-Fe is compared against \textit{ab initio} studies \cite{Heiliger05, Heiliger06, Heiliger08}. In SBTB, the current density in P and AP configuration for each spin orientation is given as:
\begin{eqnarray}J = \frac{e}{h}\int dE_l\ T_{SBTB}\ [f_{1}-f_{2}]\end{eqnarray}
The $\Delta_1$ band current density in AP configuration is nearly zero due to its half-metallic nature, thus the total AP current density is dominated by the $\Delta_5$ band. We thus, systematically extract the parameters for $\Delta_5$ band and then for the $\Delta_1$ band by matching with the AP and the P current density respectively as given by the \textit{ab initio} calculations \cite{Heiliger05,Heiliger06,Heiliger08}. $t_o$ is taken the same for both the bands and $U_b$ for the $\Delta_1$ and the $\Delta_5$ band is used as a fitting parameter. Hence, in the SBTB model for Fe-MgO-Fe MTJ there are three fitting parameters, $U_b$ for $\Delta_1$ and $\Delta_5$ bands and $t_o$. It is worthwhile to mention here that unlike the Wentzel-Kramers-Brillouin (WKB) approximation, where only the product of effective mass $m^*$ and $U_b$ matters, we find that in SBTB, only the reported values of $U_b$ and $t_o$ gave quantitative agreement with the current levels for various barrier widths. 

\section{Discussion of results}
\begin{figure}
\centering
\includegraphics[width=3.5in]{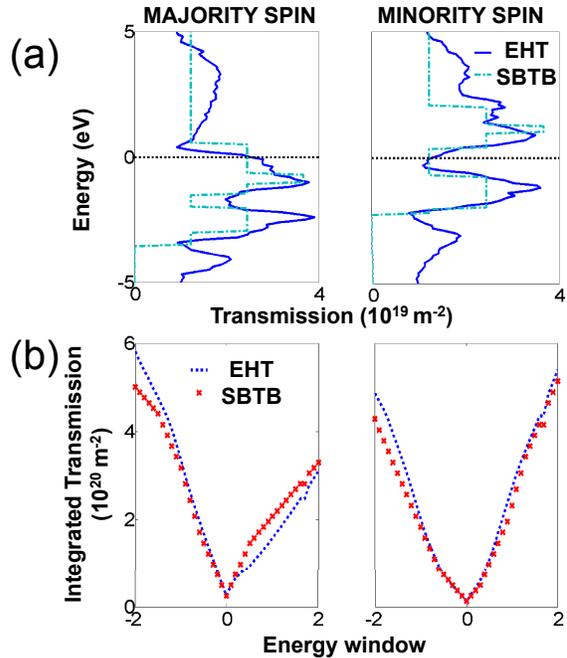}
\caption{Comparison of the energy resolved transmission and the integrated transmission through bulk Fe using SBTB and EHT for the majority and the minority spin. (a) The energy resolved transmissions using the two methods match qualitatively. The transmissions are not matched below about -3eV and -2eV for the majority and the minority spin respectively. (b) The reference for integration is taken as $E_f=0$.}
\end{figure}

In Fig. 2, the bandstructure of bcc Fe calculated using EHT is shown in the [100] direction for the majority spin for different values of $k_{||}$. In general, over the 2D transverse BZ, these bands do not remain parabolic and hence it is not possible to represent these band dispersions by suitable effective masses. Furthermore, they also have different band edge $E_{bo}$. We find a similar trend for the minority spin bands (not shown). Fig 3 is a plot of $k_{||}$ resolved equilibrium transmission at the Fermi energy using EHT for the majority and the minority spin. Here the peaks in transmission are not only present at the $\Gamma$ point, but are dispersed throughout the BZ. Therefore, it is imperative to consider the transverse BZ while calculating the transport. This is also evident in the rich transmission features as shown in Fig. 5(a) for the majority and the minority spins using EHT. These features in transmission are due to varying band offsets and bandwidths for various $k_{||}$ over the transverse BZ. 

In Fig. 5(a), the energy resolved transmission (given by Eq. 4) calculated using SBTB is compared with one calculated using EHT. In Fig. 5(b), a comparison of integrated transmission using SBTB and EHT is shown for the majority and the minority spins. Even though the energy resolved transmission using SBTB matches qualitatively with EHT calculation, the integrated transmission for various energy windows match quantitatively, and this is the most relevant quantity for the quantum transport.

\begin{figure}
\centering
\includegraphics[width=3.5in]{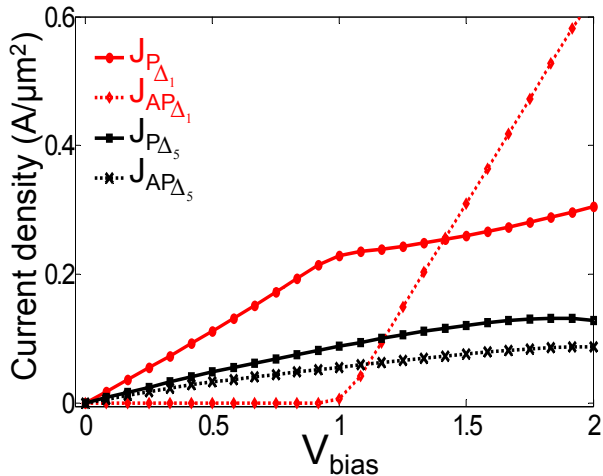}
\caption{Current density for the $\Delta_{1}$ and $\Delta_{5}$ bands for a 4-layer Fe-MgO-Fe MTJ device. Current densities for parallel and anti-parallel configurations using the SBTB method. The parameters for MgO are shown in Table II and for Fe in Table I. The low bias transport in the P configuration is dominated by the $\Delta_1$ band and in the AP configuration it is dominated by the $\Delta_5$ band.}
\end{figure}

We also compare the low bias conductance for bcc Fe(100) calculated using SBTB and EHT methods. The low bias conductance ($G_o$) at finite temperature is given as \cite{HassanSi08}:
\begin{eqnarray}G_{o}= \frac{2q^2}{hkT}\int_{\mu_o-10kT}^{\mu_o+10kT} dE T(E) \frac{e^{(E-\mu_o)/kT}}{[1+e^{(E-\mu_o)/kT}]^2}\end{eqnarray} where $\mu_o$ is the contact Fermi energy. We obtain $G_o$ as $2.16\times10^{15}\ Sm^{-2}$ and $2.23\times10^{15}\ Sm^{-2}$ for the majority and $1.08\times10^{15}\ Sm^{-2}$ and $1.14\times10^{15}\ Sm^{-2}$ for the minority spin using SBTB and EHT respectively, which match reasonably with each other. 

In Fig. 6, the calculated current densities for $\Delta_1$ and $\Delta_5$ bands using the SBTB parameters are shown for a 4-layer Fe-MgO-Fe device. These parameters are transferable for various barrier widths and a quantitative agreement in current densities and TMR ratios is obtained with \textit{ab initio} calculations \cite{Heiliger05, Heiliger06, Heiliger08} as shown in Fig. 1. $J_{P}$ is dominated by the $\Delta_1$ band current density at low bias due to a lower potential barrier seen by the $\Delta_1$ band. On the other hand, the total AP current density is dominated by the $\Delta_{5}$ band. After a certain voltage, there is a sharp increase in AP current of $\Delta_1$ band due to its half-metallic nature. Due to this increase, there is a sharp roll-off in TMR, which ultimately becomes negative. These high-bias predictions have also been reproduced in Ref. \cite{Ivan08}. The TMR calculated using the SBTB method is shown Fig. 1. TMR values match well with the ones obtained using the \textit{ab initio} model in Refs. \cite{Heiliger05, Heiliger06, Heiliger08}. To the best of our knowledge, such a bias dependence and quantitative agreement has not yet been captured within a simple model whose computational efficiency is on the order of an effective mass model.

\section{Conclusions}

We have presented a single-band tight-binding model for studying heterostructures with Fe-MgO-Fe(100) magnetic tunnel junction devices as an example due to their technological importance. We have tried to capture the band structure effects by using band parameters for various symmetry bands in the contacts and the barrier region. Features in TMR which are manifestation of the electronic structure of material were captured quantitatively within this simple model, whose computational complexity is on the order of an effective mass model. Within the same set of parameters, I-V characteristics, voltage dependence of TMR and thickness dependence of TMR were captured quantitatively. 

\section*{Acknowledgments}
We thank S. Datta for very useful discussions. The authors are also thankful to Dr. Heiliger for useful discussions and for sharing their data in electronic format. T. Z. Raza acknowledges support by the MARCO Focus Center for Materials, Structure and Devices. H. Raza is thankful to National Science Foundation (NSF) and to Nanoelectronics Research Institute (NRI) through Center for Nanoscale Systems (CNS) at Cornell University. We thank nanohub.org and NSF sponsored Network for Computational Nanotechnology (NCN) for computational resources. T. Z. Raza is also thankful to E. C. Kan for providing office space and physical resources. 
\section*{Appendix}
For bcc Fe, EHT transferable parameters are adapted from Ref. \cite{Cerda00} and are also given in Table III. These parameters are benchmarked against Slater Koster parameters \cite{Book_Papa, Slater54}. EHT is a semi-empirical tight-binding method with non-orthogonal Slater type orbitals basis set. This method has been used for large systems due to its computational simplicity \cite{Hassan07}. The overlap matrix elements are calculated as $S_{ij}=<i|j>$. The diagonal elements of Hamiltonian (H) are the ionization energies of the corresponding orbitals. The off-diagonal elements are constructed using the overlap matrix (S) as:
\begin{table}[!t]
\renewcommand{\arraystretch}{1.3}
\caption{EHT parameters for Fe majority and minority spin adapted from Ref. \cite{Cerda00}. $K_{EHT}=2.3$.}
\label{tab:Table2}
\centering
\begin{tabular}{lccccc}
\hline\hline
Orbital & $E_{on-site}(eV)$ & $c_1$ & $c_2$ & $\xi_1$ ($\AA^{-1}$) & $\xi_2$ ($\AA^{-1}$)\\
\hline\hline
$Fe^{\uparrow}$: 4s  & -9.5543 & 0.5892 &                 & 1.4884 &         \\
$Fe^{\uparrow}$: 4p  & -6.8247 & 0.5959 &                 & 1.2526 &         \\
$Fe^{\uparrow}$: 3d  & -11.9179 & 0.2396 & 0.8434        & 1.4891 &  3.3483        \\
\hline     
$Fe^{\downarrow}$: 4s  & -9.6938  & 0.5892  &         & 1.4884 &         \\
$Fe^{\downarrow}$: 4p  & -7.0227  & 0.5959  &         & 1.2526 &         \\
$Fe^{\downarrow}$: 3d  & -10.3697 & 0.3229  & 0.7796 &1.4891  & 3.3483       \\
\hline
\end{tabular}
\end{table}
\begin{eqnarray}H_{ij} = \frac{1}{2}KS_{ij}(H_{ii}+H_{jj})\end{eqnarray}
The real space Hamiltonian $H(\vec{r})$ and the overlap $S(\vec{r})$ matrices are transformed into Fourier ($\vec{k}$) space as :
\begin{eqnarray}H(\vec{k}) = \sum_{m=1}^N H_{mn}e^{i\vec{k}.(\vec{r}_{m}-\vec{r}_{n})}\end{eqnarray}
\begin{eqnarray}S(\vec{k}) = \sum_{m=1}^N S_{mn}e^{i\vec{k}.(\vec{r}_{m}-\vec{r}_{n})}\end{eqnarray}
where $\vec{k}=(\vec{k_{x}}, \vec{k_{y}}, \vec{k_{z}})$. The integer n represents the center unit cell and m represents the neighboring unit cells. 
The band structure for the majority spin channel is shown in Fig. 2 for various transverse wave vectors. Next, the equilibrium transmission is calculated for a ten layer device (five unit cells) with an infinite cross-section. For calculating the transmission, we use the equilibrium scheme of NEGF formalism \cite{Datta_Book}. This calculation was also checked by a more simple numerical calculation by counting the independent propagating modes at a particular energy for bcc Fe. The infinite device cross-section area allows to transform the real space Hamiltonian and overlap matrices in the transverse direction to reciprocal space. Thus, for each transverse reciprocal $\vec{k}_{||}=(\vec{k_y},\vec{k_z})$, we have a one dimensional (1D) lattice. Finally, the energy resolved transmission per unit area is obtained by summation over $\vec{k_{||}}$ as:
\begin{eqnarray}T_{EHT}(E)=\frac{1}{A}\sum_{\vec{k_{||}}}\tilde{T}(\vec{k_{||}})=\frac{1}{4\pi ^2}\int d\vec{k_{||}}\tilde{T}(\vec{k_{||}})\end{eqnarray}
and is shown in Fig. 4 for the majority and the minority spin channels, using $441$ $k_{||}$ points.


\begin{thebibliography}{100}
\bibitem{Magnetic_Book} H. Zabel and S. D Bader, \textit{Magnetic Heterostructures: Advances and Perspectives in Spinstructures and Spintransport} (Springer, Berlin, Heidelberg, 2008).
\bibitem{Cerda00} J. Cerda and F. Soria, "Accurate and transferable extended H\"uckel-type tight-binding parameters," \textit{Phys. Rev. B}, vol. 61, pp. 7965, 2000.
\bibitem{Heiliger05} C. Heiliger, P. Zahn, B. Y. Yavorsky and I. Mertig, "Influence of the interface structure on the bias dependence of tunneling magnetoresistance," \textit{Phys. Rev. B}, vol. 72, pp. 180406(R), 2005.
\bibitem{Heiliger06} C. Heiliger, P. Zahn, B. Y. Yavorsky and I. Mertig, "Interface structure and bias dependence of Fe/MgO/Fe tunnel junctions: \textit{Ab initio} calculations," \textit{Phys. Rev. B}, vol. 73, pp. 214441, 2006.
\bibitem{Heiliger08} C. Heiliger, P. Zahn, B. Yu. Yavorsky and I. Mertig, "Thickness dependence of the tunneling current in the coherent limit of transport," \textit{Phys. Rev. B}, vol. 77, pp. 224407, 2008. 
\bibitem{Martin04} R. M. Martin, \textit{Electronic Structure: Basic Theory and Practical Methods} (Cambridge University Press, Cambridge, UK, 2004).
\bibitem{Slater54} J. C. Slater and G. F. Koster, "Simplified LCAO Method for the Periodic Potential Problem," \textit{Phys. Rev.}, vol. 94, pp. 1498, 1954.
\bibitem{HassanSi} H. Raza, K. H. Bevan, D. Kienle, "Incoherent transport through molecules on silicon in the vicinity of a dangling bond," \textit{Phys. Rev. B}, vol. 77, pp. 035432, 2008.
\bibitem{HassanacGNR} H. Raza, E. C. Kan, "Armchair graphene nanoribbons: Electronic structure and electric field modulation," \textit{Phys. Rev. B}, vol. 77, pp. 245434, 2008.
\bibitem{Ikeda07} S. Ikeda, J. Hayakawa, Y. M. Lee, F. Matsukura, Y. Ohno, T. Hanyu and H. Ohno, "Magnetic Tunnel Junctions for spintronic memories and beyond," \textit{IEEE Transactions on electron devices}, vol. 54, No. 5, pp. 991, May 2007.
\bibitem{Butler01} W. H. Butler, X.-G. Zhang, T. C. Schulthess and J. M. MacLaren, "Spin dependent tunneling conductance of Fe/MgO/Fe sandwiches," \textit{Phys. Rev. B, Condens. Matter}, vol. 63, no. 5, p. 054416, Feb. 2001.
\bibitem{Mathon01} J. Mathon, and A. Umerski, "Theory of tunneling magnetoresistance of an epitaxial Fe/MgO/Fe(001) junction," \textit{Phys. Rev. B, Condens. Matter}, vol. 63, no. 22, p. 220403, Jun. 2001.
\bibitem{Yuasa04} S. Yuasa, T. Nagahama, A. Fukushima, Y. Suzuki and K. Ando, "Giant room-temperature magnetoresistance in single-crystal Fe/MgO/Fe magnetic tunnel junctions," \textit{Nature Materials}, vol. 3, no. 12, pp. 868-871, Dec. 2004.
\bibitem{Parkin04} S. S. P. Parkin, C. Kaiser, A. Panchula, P. M. Rice, B. Hughes, M. Samant and S.-H. Yang, "Giant tunneling magnetoresistance at room temperature with MgO(100) tunnel barriers," \textit{Nature Materials}, vol. 3, no. 12, pp. 662-867, Dec. 2004.
\bibitem{Zhang04} C. Zhang, X. -G. Zhang, P. S. Krstic, H. -P. Cheng, W. H. Butler and J. M. MacLaren, "Elecronic structure and spin-dependent tunneling conductance under a finite bias," \textit{Phys. Rev. B}, vol. 69, pp. 134406, April 2004. 
\bibitem{Derek07} D.Waldron, L. Liu and H. Guo, "Ab initio simulation of magnetic tunnel junctions," \textit{Nanotechnology}, vol. 18, pp. 424026, Sept. 2007.
\bibitem{Ivan08} I. Rungger, O. N. Mryasov and S. Sanvito, "Resonant electronic states and I-V curves of Fe/MgO/Fe(100) tunnel junctions," \textit{arXiv:0808.0902}.
\bibitem{Kittel_Book} C. Kittel, \textit{Introduction to Solid State Physics} (Wiley, New York, ed. 7, 1996).
\bibitem{Datta_Book} S. Datta, \textit{Quantum Transport: Atom to Transistor} (Cambridge University Press, Cambridge, UK, 2005).
\bibitem{Tiusan07} C. Tiusan, f. Greullet, M. Hehn, F. Montaigne, S. Andrieu and A. Schuhl, "Spin tunneling phenomena in single-crystal magnetic tunnel junction systems," \textit{J. Phys. Condens. Matter}, vol. 19, pp. 165201, 2007.
\bibitem{HassanSi08} H. Raza, T. Z. Raza, E. C. Kan, "Electrical transport in a two-dimensional electron and hole gas on a $Si(001)-(2\times1)$ surface," \textit{Phys. Rev. B}, vol. 78, No. 19, pp. 194401, 2008.
\bibitem{Book_Papa} D. A. Papaconstantopoulos, \textit{Handbook of Band Structure of Elemental Solids} (Plenum, New York, 1986).  
\bibitem{Hassan07} H. Raza, "Theoretical study of isolated dangling bonds, dangling bond wires, and dangling bond clusters on a $H:Si(001)-(2\times1)$ surface," \textit{Phys. Rev. B}, vol. 76, pp. 045308, 2007. 
\end{thebibliography}
\end{document}